\pdfoutput=1

\PassOptionsToPackage{table}{xcolor}
\documentclass{egpubl}
\usepackage{egsr2026_DLOnlyTrack}
 
\WsPaper           

\usepackage[T1]{fontenc}
\usepackage{dfadobe}  

\usepackage{cite}  
\BibtexOrBiblatex
\electronicVersion
\PrintedOrElectronic
\usepackage{graphicx} \pdfcompresslevel=9

\usepackage{egweblnk} 

\usepackage{xcolor}   

\newif\ifshowchanges
\showchangestrue        

\title[Rendering 3D Gaussians on a Graph Processor]%
      {Rendering 3D Gaussians on a Graph Processor}

        \author[N.\ Fry et al.]
{
Nicholas Fry\orcid{0009-0000-1251-9297}\textsuperscript{1},
Ignacio Alzugaray\orcid{0000-0002-7121-0000}\textsuperscript{1},
Mark Pupilli\textsuperscript{2},
Paul H. J. Kelly\orcid{0000-0001-5905-1804}\textsuperscript{1},
Andrew J. Davison\orcid{0000-0002-3784-099X}\textsuperscript{1}
\\
\textsuperscript{1}Imperial College London, Department of Computing\\
\textsuperscript{2}Independent Researcher
}

\usepackage{booktabs}
\usepackage{amsmath}
\usepackage{amssymb}
\usepackage{todonotes}
\usepackage{subcaption}
\usepackage{colortbl}
\usepackage{graphicx}
\usepackage{cleveref}

\begin{document}

\teaser{
   \vspace{-6mm}

  \includegraphics[width=\linewidth]{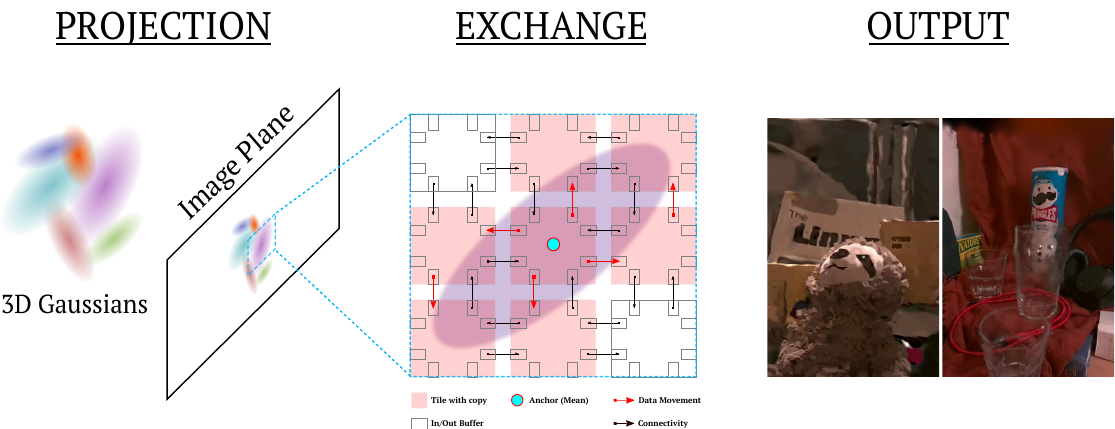}
    \vspace{1mm}
  
  \caption{3D Gaussian rendering on an Intelligence Processing Unit (IPU). \textbf{Left:} 3D Gaussians are projected onto the image plane following the standard 3D Gaussian Splatting formulation. \textbf{Centre:} the framebuffer is partitioned across IPU tiles, each of which `owns' a rectangular screen-space region held in local SRAM; projected Gaussians are routed to their anchor tile via hops on a Manhattan grid, then propagated to neighbouring tiles whose regions they overlap. Each tile independently sorts and alpha-blends its locally stored Gaussians in parallel, with no access to global memory. \textbf{Right:} rendered output on real-world scenes.}
  \label{fig:teaser}
   \vspace{6mm}
}

\maketitle

\begin{abstract}

We present the first implementation of a 3D Gaussian renderer on Graphcore's Intelligence Processing Unit (IPU), comprising 1{,}472 independent tiles with only on-chip SRAM; constraints that approximate key properties of efficient sensor-processor architectures. Our input scenes are 3D Gaussian maps from real-world sequences. Each tile `owns' a screen-space region of the framebuffer; Gaussian primitives are routed to destination tiles via Manhattan-distance hops on a north-east-west-south (NEWS) grid, then distributed to overlapping neighbours in an expanding tree pattern. Computation follows the IPU's Bulk Synchronous Parallel (BSP) model, with inter-tile communication defined at compile time. We show this hardware allows us to exploit data locality in a way that is impossible on most GPU architectures.  We evaluate the bottlenecks in this SRAM-only implementation: inter-tile bandwidth, per-tile SRAM capacity, and workload imbalance from non-uniform Gaussian density. We analyse how these constraints affect performance and render quality. This exploration raises broader questions for conventional GPUs and 3D representations, suggesting that direct inter-SM (streaming multiprocessor) communication might offer ways to reduce DRAM access in GPU kernels. We discuss these implications for the future of on-sensor and DRAM-free architectures.


\end{abstract}

\section{Introduction}
\label{sec:introduction}

3D Gaussian Splatting (3DGS)~\cite{kerbl20233d} has become the
leading method for real-time novel view synthesis. By modelling scenes
as collections of anisotropic 3D Gaussians that are projected and
alpha-composited in screen space, 3DGS achieves photorealistic
rendering quality while remaining fully differentiable. However, most
3DGS implementations assume GPU hardware with large off-chip DRAM, high
memory bandwidth, and a global address space accessible by all
threads. 3D Gaussian rendering is memory-bound on GPU: DRAM access, not
arithmetic, dominates frame time. Gaussians are loaded into shared
memory within thread blocks at the start of every kernel, so each stage
repeatedly moves data through the memory hierarchy~\cite{lee2024gscore}.
The algorithm relies on global memory, so any thread may fetch any
Gaussian on demand, the programmer can therefore treat data movement between cores as a hardware concern rather than an algorithmic one.

Efficient computing relies on memory locality: minimising the
distance between data and compute reduces latency and energy.  Computer graphics, meanwhile, concerns the relationship
between a 3D scene and the image plane, which exhibits inherent spatial
and temporal locality as the camera moves through space. Since data movement is often the dominant cost on modern hardware, an efficient rendering system should exploit this rather than fetch data each frame.
Rendering is parallel in screen space; however, 3DGS involves
\emph{global} data movement: any Gaussian can project to any screen
location, and Gaussians with large spatial extent must be shared across
multiple screen-space regions. On a GPU this is handled by global
random-access memory, at the cost of memory latency and power. On an architecture without random-access memory, this data movement must be made explicit.

`On-sensor' computing demonstrates how exploiting data locality with
pixel processor arrays (PPAs) lets meaningful computer-vision tasks run entirely on-chip with extremely constrained resources. This motivates a question for rendering: if the front-end of a spatial computing pipeline (feature detection, tracking, mapping) can run on massively parallel processors with only local memory, what does rendering look like on similar hardware? There is growing interest in deploying 3DGS on this kind of alternative hardware, driven by the desire to perform real-time differentiable rendering on resource-constrained platforms. Online 3DGS is attractive, as the same
representation serves both geometry (for tracking, mapping, and planning) and photorealistic rendering, making it well suited to AR glasses, mobile robots, and teleoperation where a reconstruction must be usable as soon as it is captured~\cite{matsuki2024gaussian, wang2024endogslam}. In this work we show that 3DGS rendering can be adapted to better exploit data locality in these settings by implementing the pipeline on a locally-connected, DRAM-free processor.

We investigate this question using a Graphcore Mk2 Intelligence Processing Unit (IPU)~\cite{graphcore_ipu}, a massively parallel processor comprising 1{,}472 independent tiles, each with 624\,KB of SRAM and no external DRAM. The IPU shares key characteristics of efficient sensor-processor architectures: local memory only, explicit inter-tile communication, and no shared address space. However, it remains programmable enough to prototype a full rendering pipeline. Our aim is not to argue that the IPU is a superior architecture for rendering, but to use its unique characteristics as a platform for experimentation: by using explicit data movement between tiles, the architecture enables data patterns that could improve the original method and inspire new representations and hardware that exploit locality to more effectively reduce power and latency.
\vspace{-3mm}

\subsection*{Contributions.}
\begin{itemize}
  \item The first implementation of 3DGS rendering on an SRAM-only, MIMD processor architecture.
  \item A routing scheme for distributing Gaussian primitives across tiles that own screen-space regions, operating within the constraints of compile-time-defined communication.
  \item Experimental analysis of rendering quality and performance of our method.
  \item Evaluation and analysis of bottlenecks: inter-tile bandwidth saturation, per-tile SRAM pressure, and load imbalance.
  \item Insights on 3D representation and algorithm design for GPUs and future DRAM-free and on-sensor architectures.
\end{itemize}
\vspace{-3mm}

\section{Related Work}
\label{sec:related}

\paragraph*{3DGS accelerators.}
Several works target 3DGS on alternative hardware. GSCore~\cite{lee2024gscore} proposes an FPGA accelerator with a custom memory hierarchy for 3DGS rasterisation, while FAMERS~\cite{famers2024}, GauRast~\cite{gaurast2024} and GCC \cite{gccaccelerator} explore ASIC designs optimised for throughput and energy efficiency. Neo~\cite{neo2026} targets on-device deployment with hardware-aware kernel design. These designs retain external memory and inherit the assumption that Gaussian data is globally accessible, optimising to minimise DRAM traffic; we take this a step further and explore whether external memory can be eliminated altogether.

\paragraph*{Incremental and online 3DGS.}
Many works in online reconstruction produce a 3D Gaussian map as the target representation~\cite{matsuki2024gaussian, Meuleman_2025_onthefly, keetha2024splatam}, optimising the map incrementally from RGB-D or monocular input as the images are received from the camera. These are still far from being hard real-time (30fps) and could benefit greatly from both hardware and algorithmic optimisations to enable the capabilities mentioned in \Cref{sec:introduction}. 

\paragraph*{Rendering on the IPU.}
Pupilli~\cite{pupilli2023neural} implemented neural path tracing entirely in SRAM on a Bow-Pod-16, with scene BVH and neural environment weights stored on-chip. That work streams ray data from a host CPU and does not dynamically redistribute scene data, it instead copies the whole scene to every tile; a key limitation we address. Recently, Tuya et al. \cite{tuya2026radiantfoamrenderinggraph} show that it is possible to also render Radiant Foam \cite{govindarajan2025radiant} on the same architecture. We examine a similar problem, but with a more widely adopted representation.

\paragraph*{On-sensor computing.}
The SCAMP architecture~\cite{dudek2005scamp} integrates processing with every pixel of an image sensor. Recent work by Bose et al.~\cite{bose2025dip} has shown that point-feature tracking can run on such hardware at thousands of frames per second on approximately 1\,W, earning a best paper nomination at CVPR 2025. Compelling examples of the on-sensor paradigm \cite{Bose_2017_ICCV, Bose_2019_ICCV_camcnn, Bose_gaze_tracking, riku_bitvo, So_2024_CVPR} establish that meaningful vision tasks should be run entirely on-sensor and motivate the present investigation into more complex on-sensor algorithms such as differentiable rendering.

\section{Background}
\label{sec:background}

\subsection{3D Gaussian Splatting}
\label{sec:bg_3dgs}

3DGS~\cite{kerbl20233d} represents a scene as a set of 3D Gaussian primitives, each parameterised by a mean $\boldsymbol{\mu} \in \mathbb{R}^3$, a 3D covariance matrix $\boldsymbol{\Sigma}$, an opacity $\alpha$, and view-dependent colour encoded via spherical harmonics. The covariance is reparameterised as $\boldsymbol{\Sigma} = R S S^T R^T$ where $R$ is a rotation matrix (stored as a quaternion) and $S$ is a diagonal scaling matrix, enabling independent optimisation of shape parameters.

Rendering proceeds in four stages: (1)~frustum culling removes Gaussians outside the view; (2)~surviving Gaussians are projected to 2D screen space, producing a 2D mean and covariance (the ``conic'' matrix $Q$); (3)~Gaussians are sorted by depth; and (4)~per-pixel alpha compositing accumulates colour contributions in front-to-back order using the EWA splatting formulation~\cite{ewasurfacesplat}:
\begin{equation}
  C(\mathbf{x}) = \sum_{i=1}^{N} c_i \, \sigma(\alpha_i) \exp\!\bigl(-\tfrac{1}{2}\bar{\mathbf{x}}_i^T Q_i \, \bar{\mathbf{x}}_i\bigr) \prod_{j=1}^{i-1}\bigl(1 - \sigma(\alpha_j) \exp\!\bigl(-\tfrac{1}{2}\bar{\mathbf{x}}_j^T Q_j \, \bar{\mathbf{x}}_j\bigr)\bigr)
  \label{eq:splatting}
\end{equation}
where $\bar{\mathbf{x}}_i = \mathbf{x} - \boldsymbol{\mu}_i^{2D}$ is the offset from the projected mean. The original GPU implementation performs a global radix sort of the Gaussians in DRAM and uses CUDA shared memory within thread blocks for tile-based rasterisation. Crucially, every stage after culling relies on either global memory or shared memory, neither of which exists on the IPU.

\subsection{The Graphcore IPU}
\label{sec:bg_ipu}

The Graphcore Mk2 IPU (GC200) is a massively parallel processor designed for machine learning workloads. It comprises 1{,}472 independent tiles, each containing a processor core with 6 hardware threads and 624\,KB of local SRAM. There is no external DRAM, no L2 cache, and no shared address space: each tile can only access its own SRAM directly. 

Execution follows the Bulk Synchronous Parallel (BSP) model. Each `superstep' consists of a \emph{compute phase} (tiles execute independently on local data), followed by an \emph{exchange phase} (tiles communicate via a structured fabric), and a global \emph{barrier synchronisation}. Communication patterns are defined at compile time through the Poplar graph compiler~\cite{poplar}: the programmer specifies a computational graph of \emph{Variables} (tensors mapped to tile memory) and \emph{Vertices} (compute kernels, called \emph{codelets}) which the compiler maps to hardware.

The key contrast with GPUs is that all data movement is explicit and predetermined. A GPU thread can load any address in global memory at runtime; an IPU tile can only send and receive tensors defined when the program is compiled. This constraint fundamentally changes how algorithms must be structured.

\subsection{Pixel Processor Arrays}
\label{sec:bg_ppa}

PPAs such as SCAMP integrate a simple ALU and local memory with every pixel. Communication is restricted to north-east-west-south (NEWS) neighbours, and all pixels execute the same instruction (SIMD). The IPU shares structural properties with PPAs: local-only memory, explicit predetermined communication, and massively parallel execution. However, IPU tiles are far more capable (624\,KB SRAM, 6 threads, full C++ programmability) compared to PPA pixels (bytes of storage, basic programmable logic). We use the IPU to explore how spatial locality can be exploited in rendering, analogous to how PPAs operate. The differences are summarised in \Cref{tab:arch_comparison}. DRAM traffic dominates in most GPU algorithms, which further motivates algorithms that avoid or limit its access.

\begin{table*}[t]
  \centering
  \caption{Architectural properties of the three classes of hardware considered in this work.}
  \label{tab:arch_comparison}
  \small
  \renewcommand{\arraystretch}{1.2}
  \rowcolors{2}{gray!8}{white}
  \begin{tabular}{@{}>{\bfseries}l c c c@{}}
    \toprule
    \rowcolor{white}
    \textbf{Property} & \textbf{GPU} & \textbf{IPU} & \textbf{PPA} \\
    \midrule
    Parallelism       & SIMT                & MIMD (BSP)             & SIMD \\
    Memory            & Global DRAM         & Local SRAM only        & Local (bytes) \\
    Communication     & Global loads        & Compile-time channels  & NEWS neighbours \\
    Per-unit memory   & $\sim$256\,KB L1    & 624\,KB SRAM           & $\sim$10\,B \\
    Programmability   & Full (CUDA)         & Full (Poplar)             & Limited (assembly) \\
    \bottomrule
  \end{tabular}
\end{table*}

\section{Method}
\label{sec:method}

\subsection{Pipeline Overview}
\label{sec:pipeline_overview}

\begin{figure}[ht]
    \centering
    \includegraphics[width=1\linewidth]{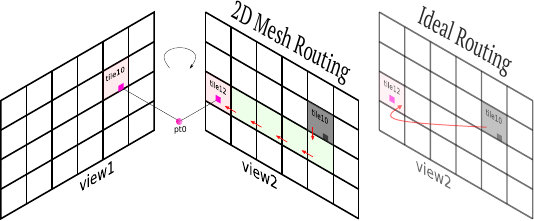}
    \caption{We show the 2D Mesh routing that is used in the IPU implementation vs the ideal routing. In our routing scheme the Gaussian migrates from tile-to-tile (shown by red arrows and green tiles). The ideal communication mechanism would enable reads from arbitrary memory locations in other tiles. The same is true in a GPU, ideally Gaussians could be routed to the correct SM without having to reload from DRAM. }
    \label{fig:networking}
\end{figure}

Our pipeline maps the four stages of 3DGS rendering onto the IPU's BSP execution model. Each BSP superstep alternates between local computation and inter-tile communication:

\begin{enumerate}
  \item \textbf{Projection} (local compute): each tile projects its stored Gaussians using the view matrix streamed from the host, computing 2D means, conics, and bounding boxes.
  \item \textbf{Routing} (compute + exchange, repeated): Gaussians whose projected means fall outside the local tile are evicted toward their destination via Manhattan-distance hops on the 2D mesh (NEWS) grid (see \Cref{fig:networking}). This phase repeats over multiple BSP supersteps until convergence.
  \item \textbf{Bloom} (compute + exchange, repeated): Gaussians that have reached their `anchor' tile (the tile which holds the part of the framebuffer the mean will be rendered on) are propagated to neighbouring tiles whose screen regions they overlap, using an expanding tree-pattern (see \Cref{fig:treeprotocol}).
  \item \textbf{Compositing} (local compute): each tile sorts its local Gaussians by depth and performs front-to-back alpha compositing onto its framebuffer region.
\end{enumerate}

\begin{figure}
    \centering
    \includegraphics[width=1\linewidth]{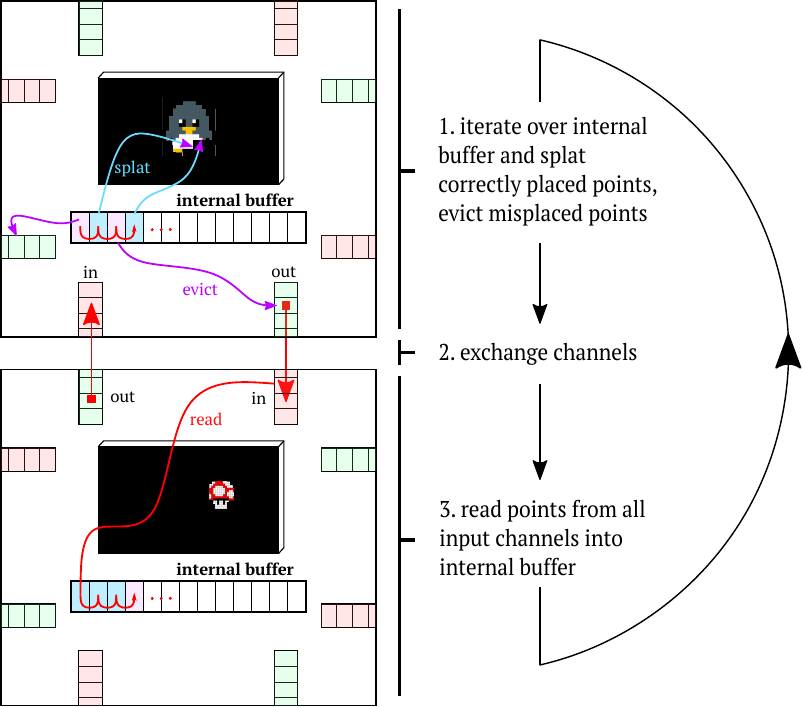}
    \caption{Kernel execution order for each tile. A Gaussian is kept if it needs be rendered, or transferred closer to its destination. Each core iterates over its z-buffer (shown as \textit{internal buffer}) and checks if the primitives' contribution should be accumulated into the framebuffer (blue arrow). If the Gaussian should be rendered on another tile, it is moved (purple arrow) to a communication channel. The tiles then exchange Gaussians, after which the new Gaussians are again either inserted in the z-buffer or evicted.  }
    \label{fig:pipeline}
\end{figure}

The only host communication per frame is the view/projection matrix (input) and the assembled framebuffer (output). \emph{Thus, all scene data remains in on-chip SRAM throughout execution}.  \Cref{fig:pipeline} shows the high-level execution of IPU kernels in the pipeline.

\subsection{Framebuffer Partitioning}
\label{sec:framebuffer}

The output framebuffer (1280$\times$720) is divided into 1{,}440 equal rectangular slices of 32$\times$20 pixels, each pinned to a separate IPU tile's SRAM. Tiles are arranged in a 2D grid matching the spatial layout of the framebuffer, so that tile adjacency corresponds to screen-space adjacency. Each tile stores its slice as 8-bit RGBX values (2.5\,KB per tile). The fourth byte pads each pixel to a 32-bit word boundary, which is required for race-free concurrent writes from multiple hardware threads on the IPU.

\subsection{Gaussian Routing via NEWS Grid}
\label{sec:routing}

The central challenge of implementing 3DGS on the IPU is distributing Gaussians to the tiles that need them, without a unified address space. On a GPU, any thread can read any Gaussian from global memory which are then allocated to the shared memory of the GPUs SMs (streaming multiprocessor) based on screenspace tile-id. In our IPU implementation, Gaussians must physically migrate through tile-to-tile channels whose connections are fixed at compile time.

\paragraph*{Topology.}
Tiles are arranged in a 2D mesh (\Cref{fig:teaser}, middle), with
fixed-size channels connecting each tile to its four cardinal neighbours
(depicted as small black rectangles on the tile borders). Separate input
and output buffers per direction allow simultaneous reads and writes,
and boundary tiles wrap their channels back to themselves. This
arrangement, inspired by Network-on-Chip routing~\cite{Chron2007RoutingAF},
guarantees that any tile can reach any other through a sequence of
single-hop transfers. The design is agnostic to tile count and maps
directly to larger grids, a property it shares with 
wafer-scale architectures~\cite{lie2023cerebras}.

\paragraph*{Destination computation.}
Each Gaussian has an \emph{anchor tile}: the tile whose screen region contains the Gaussian's projected 2D mean. After projection, a tile compares each of its Gaussians against its local screen bounds. Gaussians whose anchors should lie elsewhere are \emph{evicted}; i.e. written to the outgoing channel that minimises Manhattan distance to the destination.

\paragraph*{Routing protocol.}
During the exchange phase, each tile's outgoing buffers are copied to the corresponding incoming buffers of its neighbours. In the subsequent compute phase, tiles read their incoming channels and either store arriving Gaussians locally (if this tile is the destination) or forward them toward the correct destination. This process repeats across BSP supersteps until all Gaussians have reached their anchor tiles. Convergence is guaranteed within $\max(W, H)$ supersteps, where $W$ and $H$ are the grid dimensions.

Each Gaussian carries a unique 32-bit ID assigned at compile time. This ID prevents duplicate storage when a Gaussian passes through intermediate tiles, and setting the ID to zero signals eviction.

\subsection{Tree-Pattern `Bloom'}
\label{sec:bloom}

\begin{figure}[h]
    \centering
    \includegraphics[width=1\linewidth]{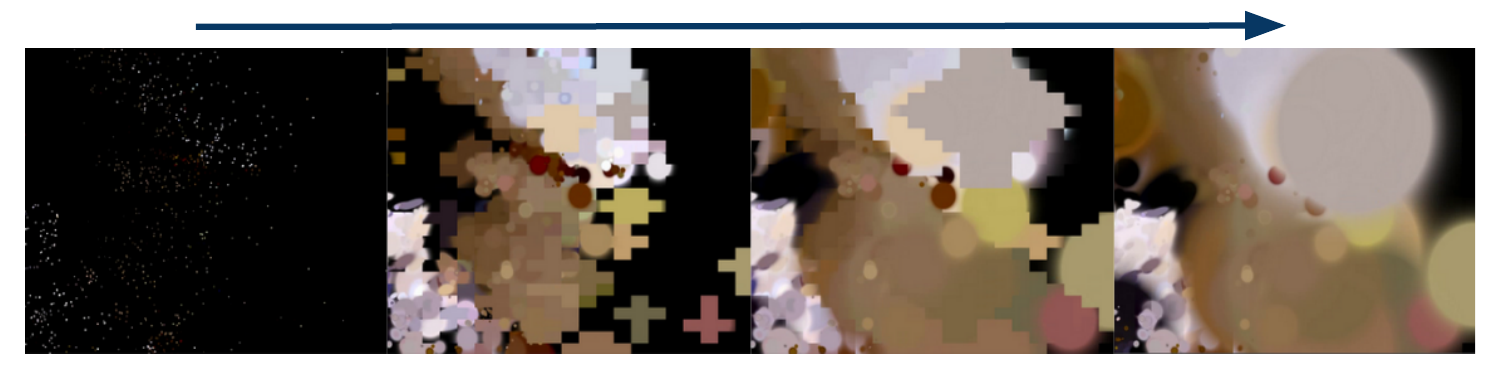}
    \caption{Gaussian `blooming' protocol. The arrow shows the order of BSP timesteps.}
    \label{fig:blooming}
\end{figure}

Once a Gaussian reaches its anchor tile, it may still need to be rendered on neighbouring tiles as its 2D screen-space extent can span multiple framebuffer tiles. The GPU implementation handles this by copying into shared memory of each SM that needs it; similarly on the IPU, we must explicitly propagate copies, but we do this using local tile-to-tile communication rather than global memory. This is shown in \Cref{fig:blooming}.

\begin{figure}[t]
    \centering
    \includegraphics[width=1\linewidth]{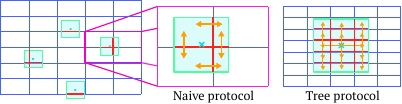}
    \caption{Gaussian `blooming' protocol based on the 2D bounding box (green). If every tile forwards (orange arrows, naive protocol) an incoming Gaussian to all neighbour tiles (blue) whose regions it overlaps (red), copies will cycle through the grid indefinitely. This means each tile will receive the same Gaussian from multiple directions, saturating inter-tile channels and occupying SRAM with duplicates. Our tree-protocol eliminates this by establishing a horizontal-then-vertical (orange arrows, tree protocol) propagation order.}
    \label{fig:treeprotocol}
    \vspace{5mm}
\end{figure}

We compute the bounding box using the same bounding-radius method as the original 3DGS: three standard deviations of the larger eigenvalue of the 2D covariance define a square bound around each Gaussian's projected mean. The bounding box determines which neighbouring tiles require a copy. Propagation follows a structured \emph{tree pattern}: the anchor tile first sends copies along horizontal beams (left/right), then tiles along these beams forward copies vertically (up/down). This ordering eliminates cyclic copying as shown in \Cref{fig:treeprotocol}.

\begin{table*}[ht]
  \centering
  \caption{Per-stage execution time (ms) by scene. Routing, projection and sorting min\,/\,mean\,/\,max across tiles. Blending is reported as wall-clock (slowest tile), as we found it difficult to accurately record the time for that codelet. The Total column sums the slowest-tile (max) contribution of each stage, since the BSP barrier is set by the slowest tile. Inter-tile exchange averages $0.07$\,ms.}
  \label{tab:timing}
  \begin{tabular}{lccccc}
    \toprule
    Scene & Blend & Route & Projection & Sort & Total$_{\max}$ \\
    \midrule
    Pringles & 17.16 & 1.48\,/\,3.72\,/\,14.11 & 3.06\,/\,3.81\,/\,11.57 & 0.00\,/\,0.24\,/\,3.75 & 46.59 \\
    Chairs   & 15.76 & 1.60\,/\,6.82\,/\,16.06 & 3.10\,/\,4.41\,/\,11.73 & 0.00\,/\,0.40\,/\,3.86 & 47.41 \\
    Salad    & 16.66 & 1.48\,/\,4.14\,/\,14.22 & 3.07\,/\,3.95\,/\,11.61 & 0.00\,/\,0.26\,/\,3.37 & 45.86 \\
    Sloth    & 16.15 & 1.48\,/\,2.71\,/\,13.70 & 3.04\,/\,3.41\,/\,11.15 & 0.00\,/\,0.12\,/\,3.40 & 44.40 \\
    \midrule
    Average  & 16.43 & 1.48\,/\,4.35\,/\,14.52 & 3.07\,/\,3.90\,/\,11.52 & 0.00\,/\,0.26\,/\,3.60 & 46.07 \\
    \bottomrule
  \end{tabular}
\end{table*}

\subsection{Rasterisation}
\label{sec:on_tile_rendering}

After routing and blooming, each tile holds a set of Gaussians that are visible within its framebuffer region. Rendering proceeds locally:

\begin{enumerate}

  \item \textbf{Culling}:  Gaussians
  that land behind the near plane or fall clearly outside the tile are
  skipped.
  
  \item \textbf{Depth sort}: The render buffer is sorted by depth using an iterative sort (recursive algorithms are not supported on IPU tiles due to dynamic stack constraints).
  
  \item \textbf{Alpha compositing}: for each pixel in the local framebuffer slice, contributions from all Gaussians in the sorted render buffer are accumulated in front-to-back order using \Cref{eq:splatting}. The per-pixel colour is computed using the EWA resampling filter, where the exponential falloff from the conic matrix provides the soft Gaussian edges.
\end{enumerate}

\paragraph*{Gaussian Representation}
\label{sec:representation}

Each Gaussian is stored as a 60-byte struct, aligned for efficient IPU memory access:

\begin{center}
\small
\begin{tabular}{ll}
  \toprule
  \textbf{Field} & \textbf{Size} \\
  \midrule
  Mean (world space, \texttt{float3}) & 12\,B \\
  Colour (RGB + opacity, \texttt{float4}) & 16\,B \\
  Rotation (quaternion, \texttt{float4}) & 16\,B \\
  Scale (log-space, \texttt{float3}) & 12\,B \\
  Gaussian ID (\texttt{uint32}) & 4\,B \\
  \midrule
  \textbf{Total} & \textbf{60\,B} \\
  \bottomrule
\end{tabular}
\end{center}

The byte alignment ensures single-cycle memory reads. View-dependent colour is simplified to a single RGB value per Gaussian (zeroth-order spherical harmonic) to reduce per-primitive memory.

\begin{figure*}[!t]
  \centering
  \setlength{\tabcolsep}{0pt}
  \renewcommand{\arraystretch}{0}
  \begin{tabular}{@{}cc@{}}
    \textbf{GPU} & \textbf{IPU (ours)} \\[2pt]
    \includegraphics[width=0.5\textwidth]{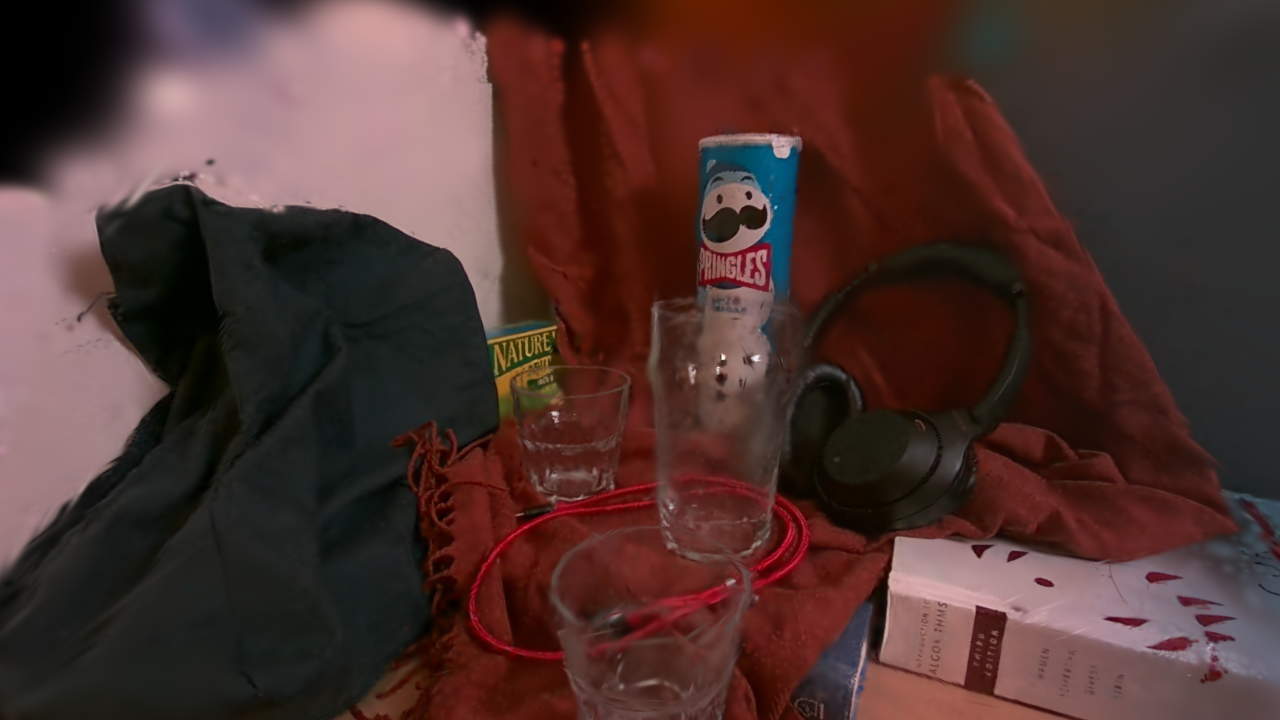} &
    \includegraphics[width=0.5\textwidth]{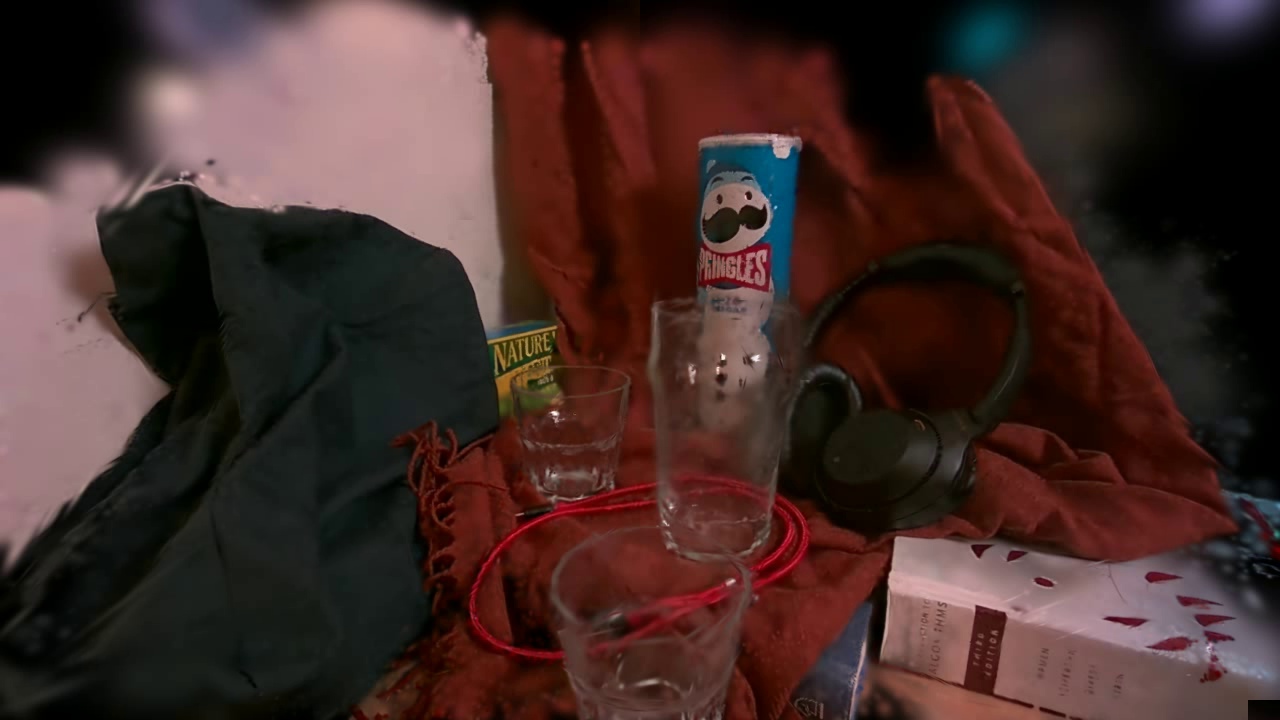} \\
    \includegraphics[width=0.5\textwidth]{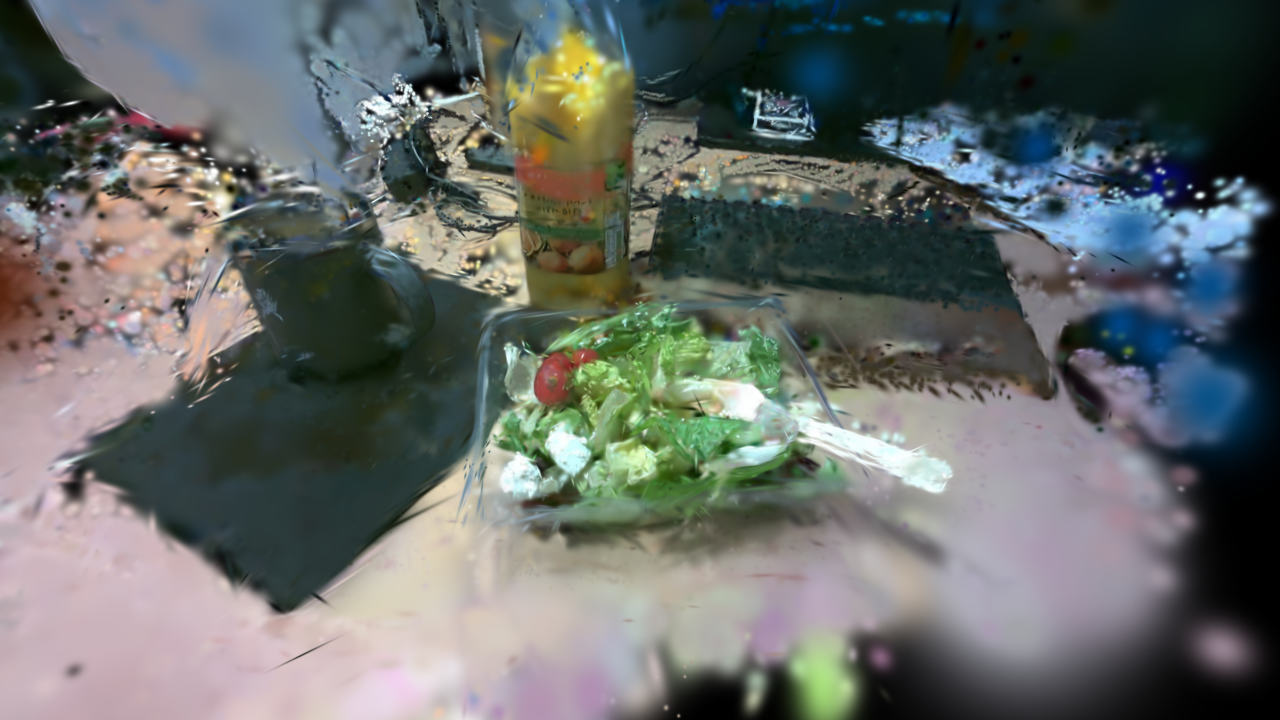} &
    \includegraphics[width=0.5\textwidth]{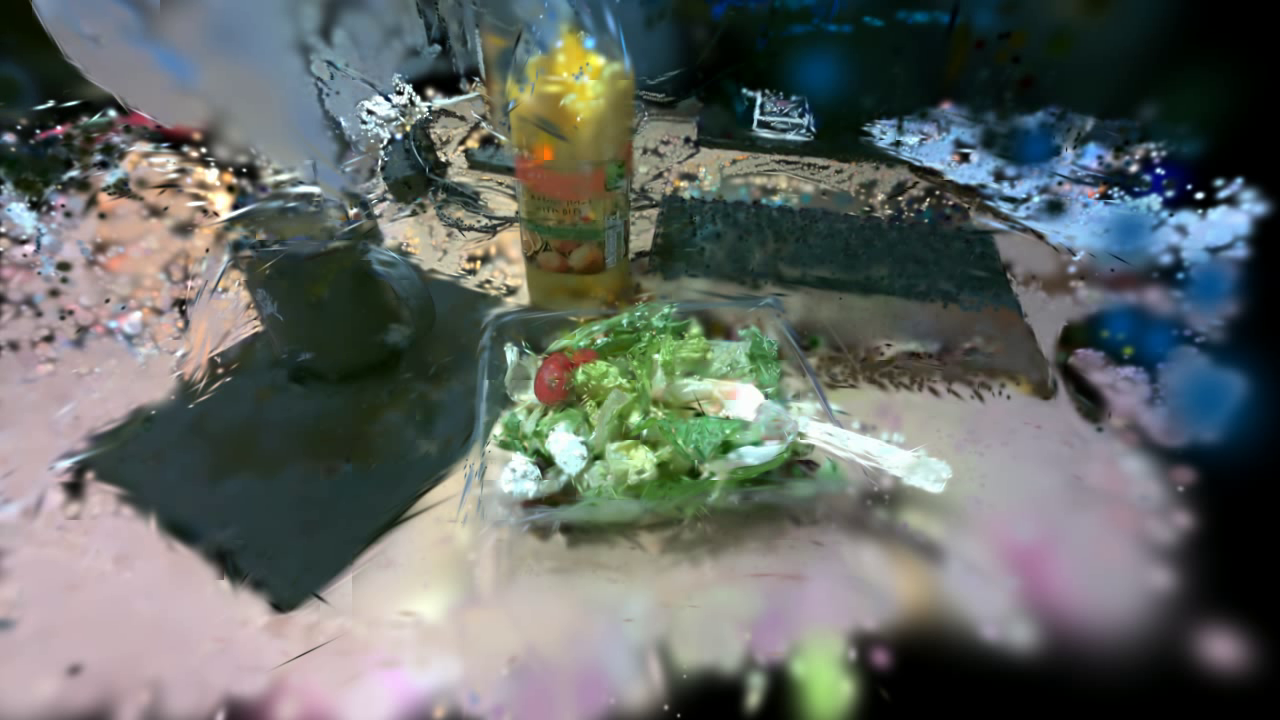} \\
    \includegraphics[width=0.5\textwidth]{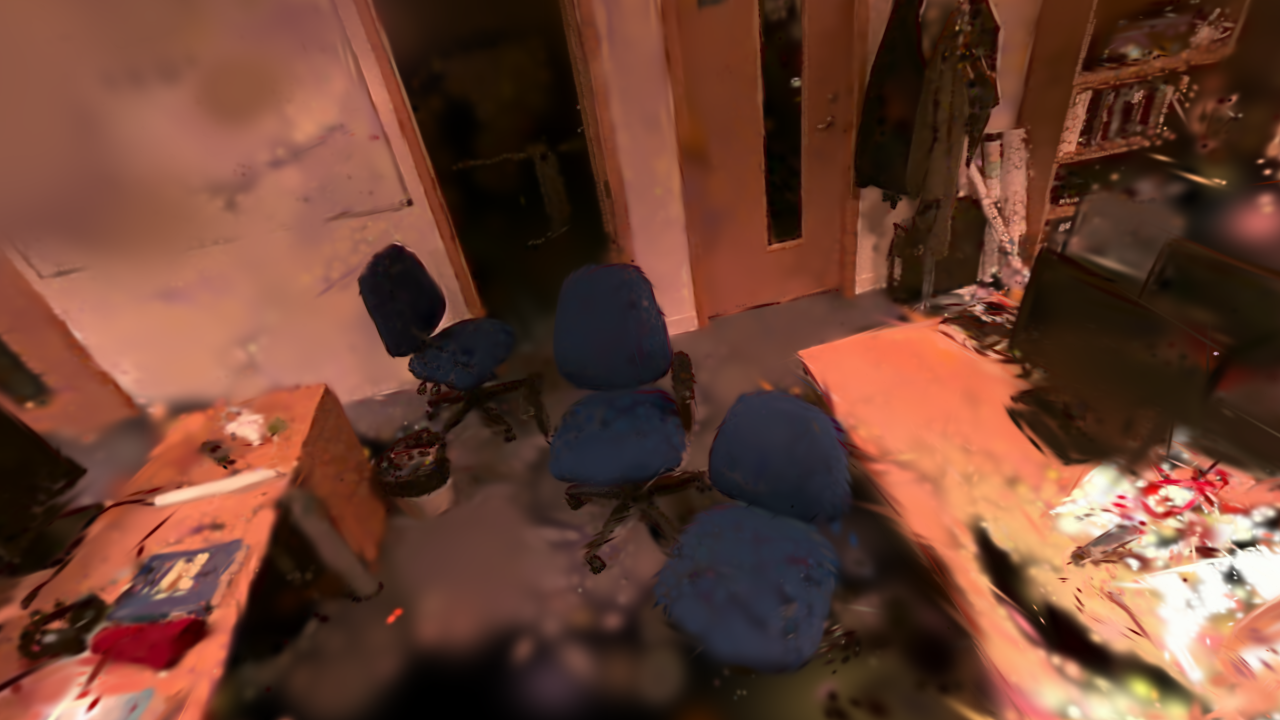} &
    \includegraphics[width=0.5\textwidth]{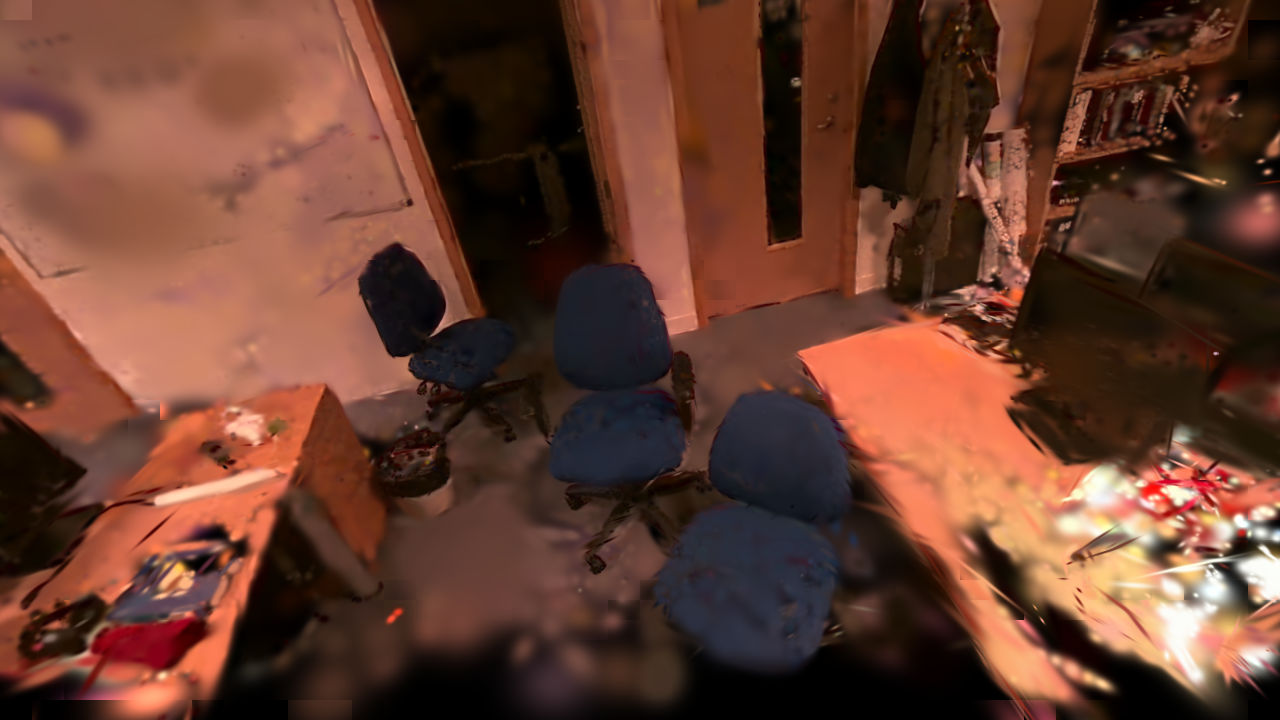} \\
    \includegraphics[width=0.5\textwidth]{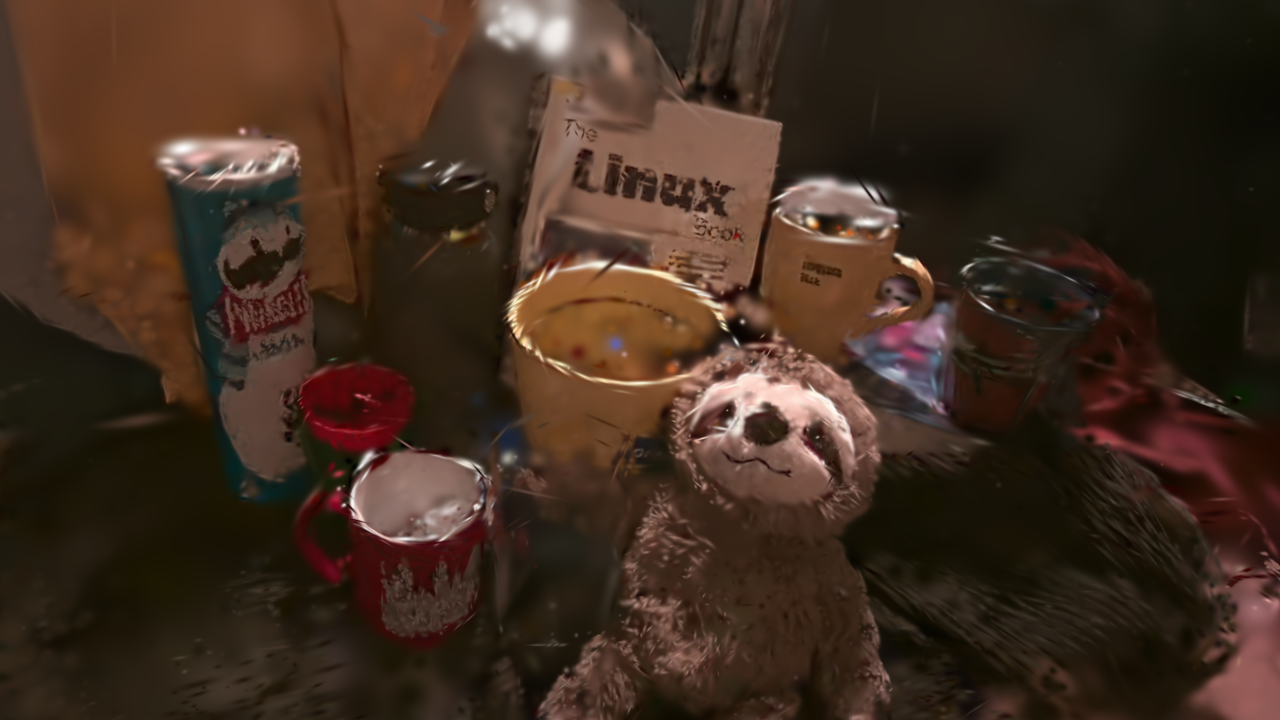} &
    \includegraphics[width=0.5\textwidth]{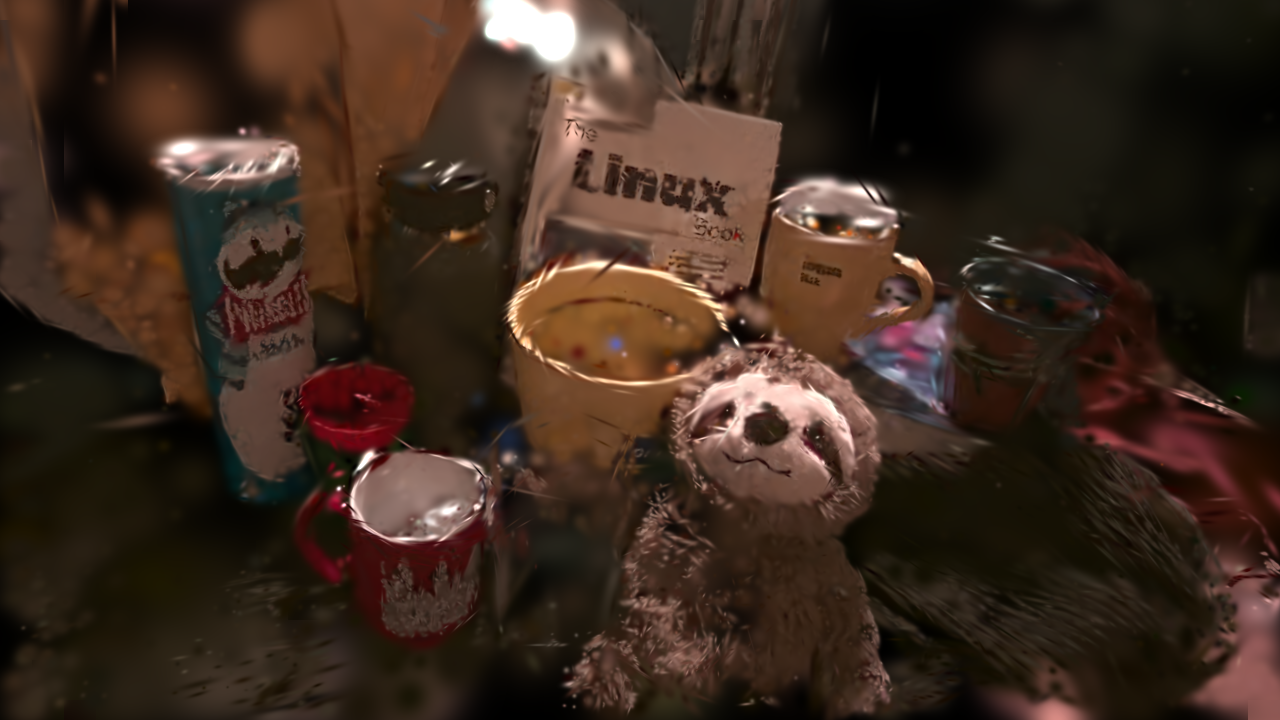} \\
  \end{tabular}
  \caption{Qualitative comparison between the GPU 3DGS baseline (left column) and our IPU implementation (right column). Rows, from top to bottom: (a)~Pringles (91{,}450 Gaussians), (b)~Salad (38K Gaussians), (c)~Chairs (44K Gaussians), and (d)~Sloth (25K Gaussians). For these low-density scenes the IPU rendering is close to the GPU baseline, with fine details such as text and thin structures preserved. }

  \label{fig:qualitative_comparison}
\end{figure*}

\section{Results}
\label{sec:results}

\subsection{Experimental Setup}

All experiments are conducted on a single Graphcore Mk2 IPU (GC200) with 1{,}472 tiles at a framebuffer resolution of 1280$\times$720. Standard 3DGS datasets would be too large to fit on-chip \cite{barron2022mipnerf360, DeepBlending2018}, so we evaluate on 3D Gaussian maps exported from Gaussian Splatting SLAM~\cite{matsuki2024gaussian} trained on real-world RGB sequences.  We test on various small scenes of different complexity: Chairs (44K Gaussians), Pringles (91K), Bonsai (273K), Salad (38K), Sloth (25K).

\subsection{Memory Budget}
\label{sec:memory}

\begin{figure}[ht]
  \centering
  \includegraphics[width=\columnwidth]{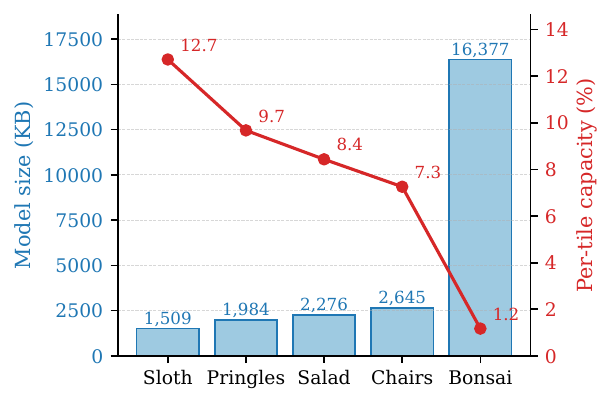}
  \caption{Model memory footprint (bars) and fraction of the model that fits in one tile's 192\,KB render buffer budget (line) across test scenes. As shown, for smaller scenes a larger percentage of the scene can fit on a single tile. }
  \label{fig:model_capacity}
\end{figure}

\Cref{fig:model_capacity} summarises the memory footprint of our
test scenes. At 60\,bytes per Gaussian, model size ranges from
1.5MB for Sloth (25{,}161 Gaussians) to 16.4MB for Bonsai
(272{,}956). Even the smallest of these is much larger than a single tile's
192KB Gaussian buffer, which can hold just 12.7\% of Sloth and
only 1.2\% of Bonsai. Copying the scene to every tile, the strategy used by
Pupilli~\cite{pupilli2023neural} for neural path tracing, is
therefore not feasible: a whole scene does not fit per tile.
Our routing scheme (\Cref{sec:routing}) addresses this
by exploiting the IPU's \emph{aggregate} memory instead:
Gaussians are distributed only to the tiles that need them, assuming a scene is evenly distributed, then the relevant capacity is 1{,}472 $\times$ 192KB rather than the ~500KB used in \cite{pupilli2023neural}. The approach is well suited to settings such as online SLAM, interactive exploration and AR where the camera moves incrementally rather than teleporting,
so each tile's working set changes only gradually between frames
and per-tile occupancy stays within budget.

\subsection{Rendered Output}

\begin{figure*}[ht]
  \centering
  \begin{subfigure}[b]{0.3333\textwidth}
    \centering
    \includegraphics[width=\linewidth]{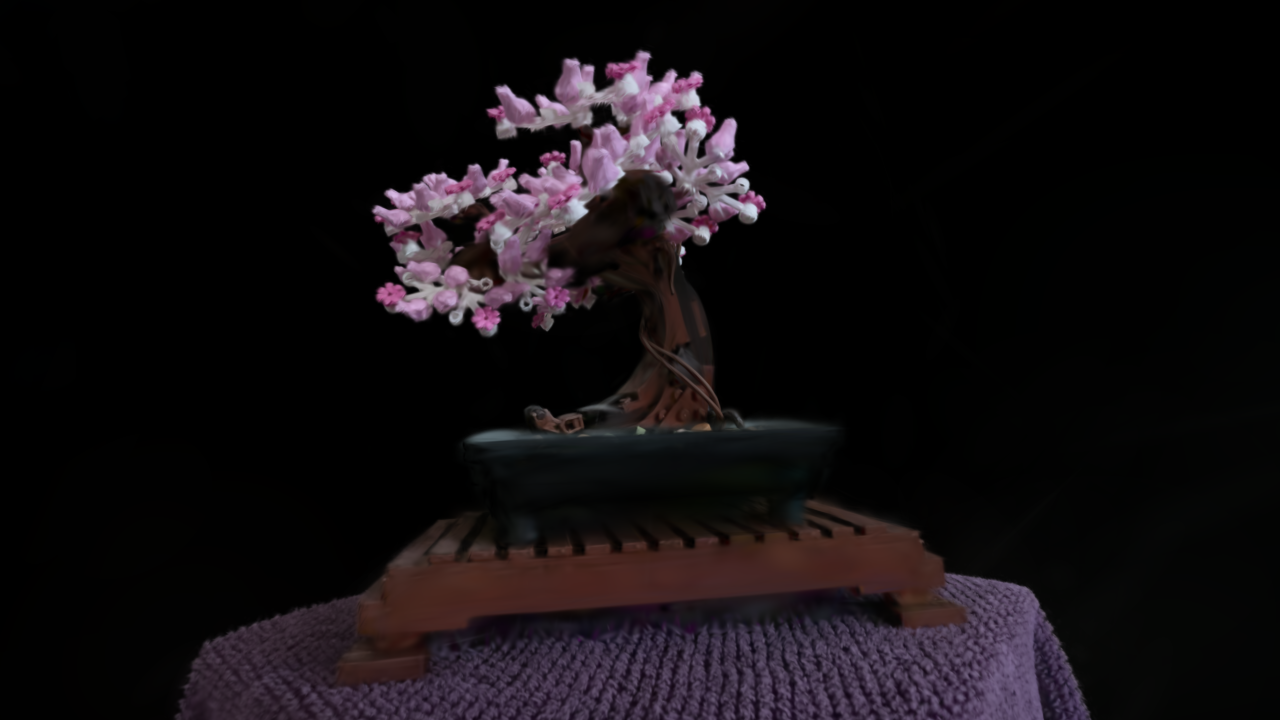}
    \caption{GPU baseline}
  \end{subfigure}%
  \begin{subfigure}[b]{0.3333\textwidth}
    \centering
    \includegraphics[width=\linewidth]{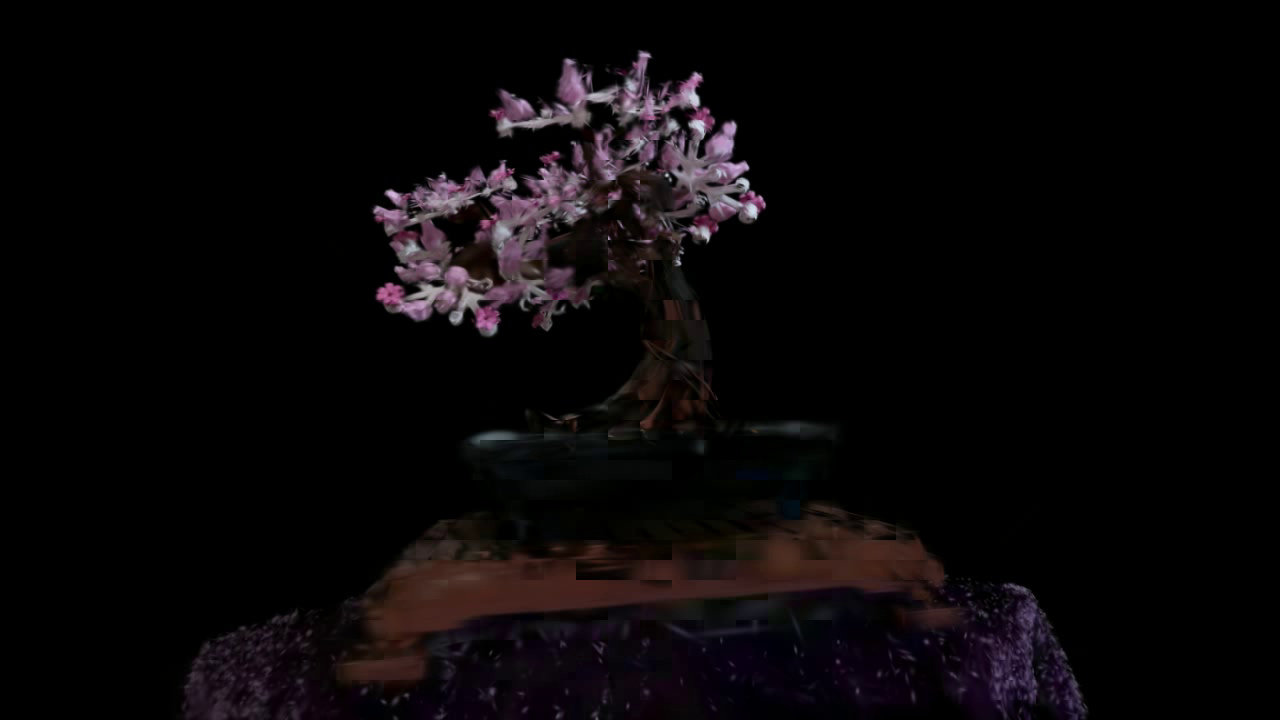}
    \caption{IPU (ours)}
  \end{subfigure}%
  \begin{subfigure}[b]{0.3333\textwidth}
    \centering
    \includegraphics[width=\linewidth]{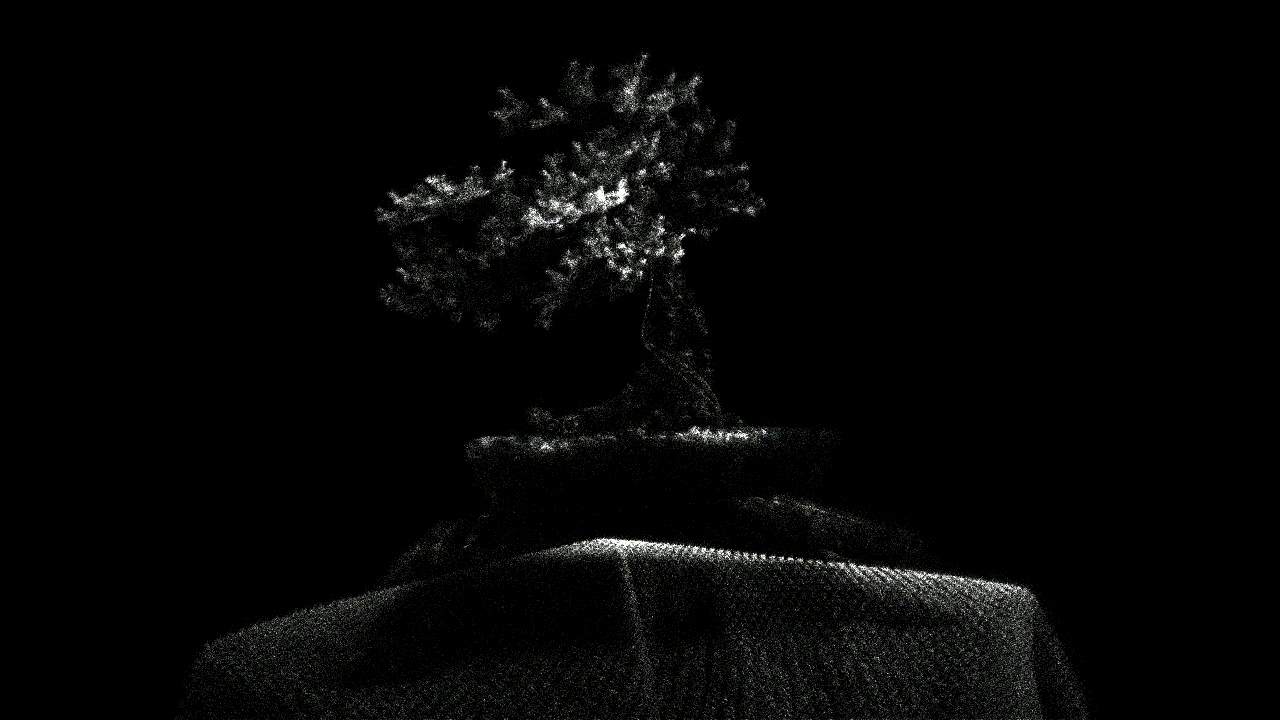}
    \caption{CPU point cloud}
  \end{subfigure}
  \caption{Bonsai scene (272{,}956 Gaussians). The IPU render captures leaf detail but exhibits tiling artifacts in dense regions where channel saturation prevents full Gaussian propagation. As the Bonsai scene is very dense in certain regions, many of the Gaussians are dropped from the rendering.}
  \label{fig:bonsai_comparison}
\end{figure*}

\begin{figure}[t]
  \centering
  \includegraphics[width=0.9\columnwidth]{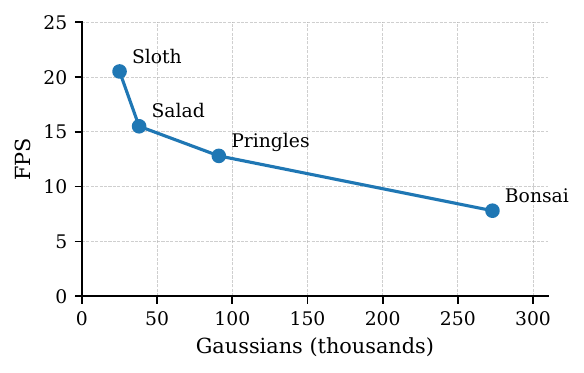}
  \caption{IPU rendering throughput vs. scene complexity at $1280{\times}720$. }
  \label{fig:fps_vs_gaussians}
\end{figure}

\begin{figure}[htbp]
  \centering
  \includegraphics[width=0.9\columnwidth]{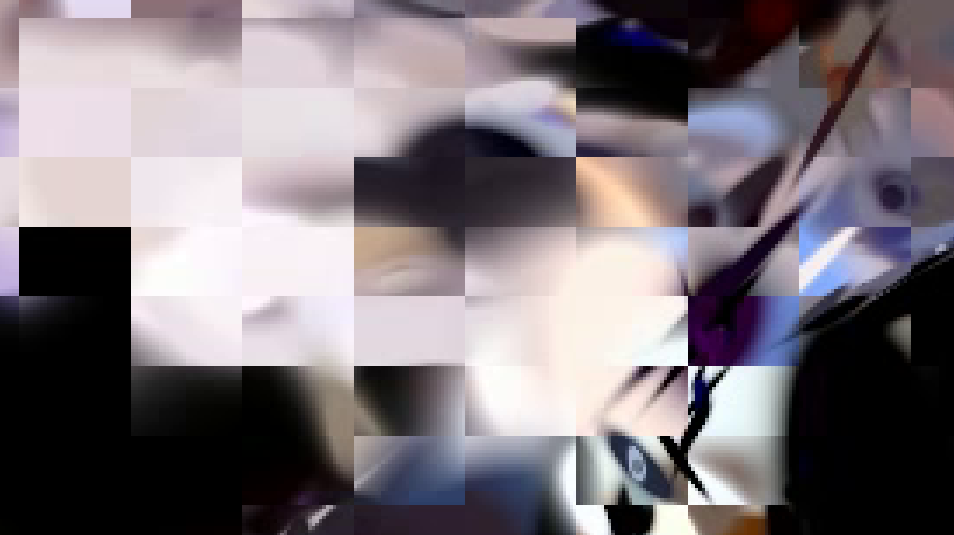}
  \caption{Tiling artifacts caused by inter-tile channel saturation. In dense regions, outgoing channels fill before all Gaussian copies can be propagated, leaving rectangular gaps aligned with tile boundaries.}
  \label{fig:tileArt}
\end{figure}

\Cref{fig:qualitative_comparison} shows rendered output compared to the GPU baseline. For these moderate-density scenes the IPU rendering is near-identical to the GPU baseline. The Pringles scene demonstrates that fine details including legible text are preserved. We show also the FPS of the scenes from these viewpoints in \Cref{fig:fps_vs_gaussians}. Despite the IPU being in no way designed for graphics workloads, the scenes render at reasonable interactive framerates. The dense Bonsai scene in \Cref{fig:bonsai_comparison} (273K Gaussians) reveals the limits of the current approach: regions of very high Gaussian density exhibit rectangular tiling artifacts (\Cref{fig:tileArt}) caused by channel saturation during the bloom phase. These artifacts are localised to areas where many Gaussians compete for limited channel capacity.

\subsection{Host Communication}

Analysis of the execution trace shows that communication between IPU SRAM and the host account for only \textbf{1.6\%} of total execution time. The remaining \textbf{98.4\%} of computation is entirely on-chip, confirming that our approach successfully avoids external memory access for scene data. This contrasts with GPU 3DGS, where DRAM bandwidth is a primary bottleneck, and with the IPU path tracer~\cite{pupilli2023neural}, which streams ray data from the host.

\begin{figure*}[t] 
    \centering
    
        \centering
        \includegraphics[width=0.99\linewidth]{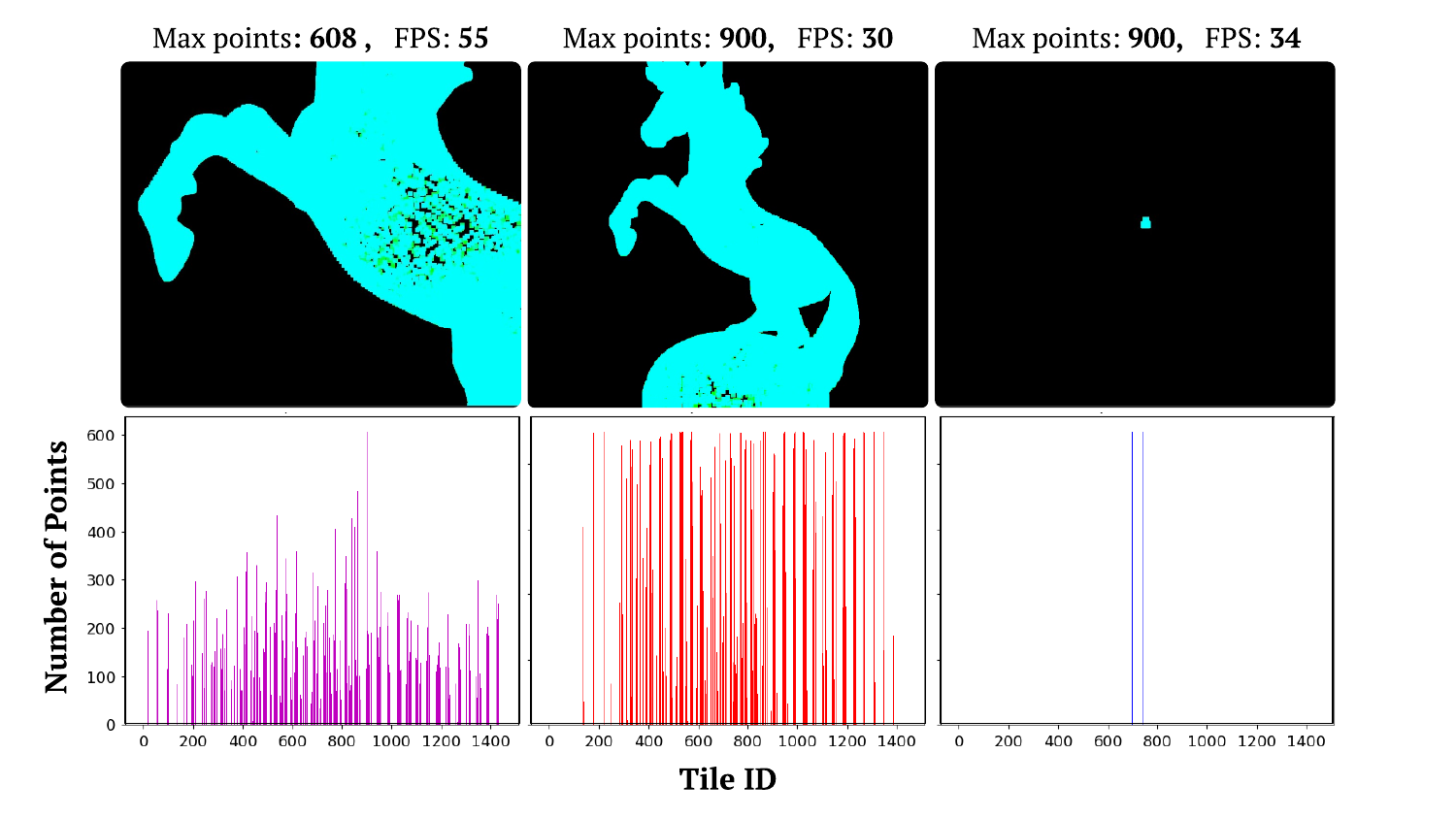}

    \caption{\textbf{Analysis of Workload Distribution.} We show three different rendered viewpoints of a model. The corresponding distribution of Gaussians across tiles is shown below. `Max points' refers to the number of points on the tile that stores the most points on the grid. For this tiny scene close-up views distribute the load more evenly across the processor, whereas distant views concentrate Gaussians on central tiles, causing imbalance. In this experiment we use a maximum render buffer size of 900 primitives. }
    \label{fig:load_balancing}
    \vspace{-3mm}
\end{figure*}

\subsection{Power Evaluation}
\label{sec:power}

We measure frame time (ms), power (W) and FPS over 1440 frames
(two full orbits of the scene) averaged across four scenes (Sloth,
Chairs, Salad, Pringles), reported in \Cref{tab:power}. An
apples-to-apples comparison is challenging as desktop GPUs are
optimised for rendering and have far more threads and complex caches. We benchmark a GTX~1080
(40,960~threads, 1.73~GHz) and RTX~4090 (262,144~threads, 2.52~GHz)
against the Mk2 IPU (8,832~threads, 1.325~GHz). The setup is identical
across chips. 

\begin{table}[t]
  \centering
  \caption{Per-frame FPS and power across different hardware, averaged over 1440 frames (two full orbits) and four scenes. The IPU draws far less power despite lower throughput.}
  \label{tab:power}
  \begin{tabular}{lrrr}
    \toprule
                    & IPU   & GTX~1080 & RTX~4090 \\
    \midrule
    Frame time (ms) & 50.55 &     1.74 &     0.42 \\
    FPS             & 19.80 &   580.55 &  2401.47 \\
    Power (W)       & 27.18 &    74.42 &    89.80 \\
    Peak power (W)  & 29.30 &   119.67 &   112.61 \\
    FPS/W           &  0.73 &     7.80 &    26.74 \\
    J/frame         & 1.373 &    0.128 &    0.037 \\
    \bottomrule
  \end{tabular}
\end{table}

\vspace{4mm}
While the IPU underperforms on FPS/W, its power draw is lower and
more stable, in a similar range to a Jetson-Orin (7--75\,W), suggesting the architecture
is suited to low-power deployment. The vendor's own benchmarks~\cite{graphcoreHotChips21}
report 571 vs 384~MFLOPS/W for a 16-chip IPU Mk2 pod versus
8$\times$~A100, attributing this to on-chip SRAM avoiding DRAM
transport, specifically $\sim$3$\times$ on data-movement.

\subsection{Timing Breakdown}
\label{sec:timing}

Using the same setup as \Cref{sec:power}, we measure the execution time of each
pipeline stage across tiles (\Cref{tab:timing}). The breakdown shows the bottleneck is alpha blending, not routing
or inter-tile bandwidth: a single exchange phase is only $0.07$\,ms,
confirming that on-chip transfer is fast. As blending is
pixel-parallel, more threads per tile would directly increase the
framerate.

\subsection{Demonstration of Data Locality}
\label{sec:locality}

To highlight the reduction in data movement, we move the camera around `Sloth' at
different rates and measure the fraction of Gaussians requiring routing
per frame (the \emph{churn-rate}). Static views on the IPU have 0\%
churn, as Gaussians do not move in memory between frames
(\Cref{tab:churn}).

\begin{table}[t]
  \centering
  \caption{Per-frame churn-rate on `Sloth' (25{,}161 Gaussians) under different camera motions. Incremental motions move only a small fraction of Gaussians between frames; only an unconstrained teleport approaches full re-routing.}
  \label{tab:churn}
  \begin{tabular}{lrr}
    \toprule
    Trajectory & Moved & Churn (\%) \\
    \midrule
    Orbit 0.1$^\circ$      &   138 &  0.55 \\
    Orbit 0.5$^\circ$      &   616 &  2.45 \\
    Orbit 2.0$^\circ$      & 2{,}745 & 10.91 \\
    Pure translation       &    56 &  0.22 \\
    Pure rotation 1$^\circ$ & 1{,}634 & 11.99 \\
    Random teleport        & 24{,}595 & 97.75 \\
    \bottomrule
  \end{tabular}
\end{table}

Churn is much lower for incremental view changes than for
teleports, confirming that data movement scales with view-change
magnitude. GPU churn is effectively 100\% per view, as all Gaussians
are potentially reloaded from DRAM (unless cached, which cannot be
guaranteed).

\subsection{Discussion}
\label{sec:discussion}

\paragraph*{DRAM-Free Constraints}

Rendering is highly parallel once each primitive is in the right place, as the alpha compositing is trivially parallel across pixels. The hard part, on both GPUs and the IPU, is getting the data to the tile that needs it. GPUs solve this with a unified global address space and a cache hierarchy: the programmer can treat any Gaussian as locally accessible, and the hardware arranges the transfers. The IPU enables a different solution: local tile-to-tile exchange, structured at compile time. What may look like a different rendering algorithm is really the same algorithm with explicit, software managed data movement. Removing the global address space makes this apparent.

Communication patterns depend on scene content, viewpoint, and Gaussian spatial extent. We show that with sudden view changes the dominant cost \emph{is} routing; moving Gaussians to the tiles that need them. However, our design allows the algorithm to exploit incremental view changes. We believe the same observation is relevant for algorithm design on a GPU. If routing is the bottleneck, then a GPU with direct inter-SM shared-memory exchange  (rather than forcing every cross-SM data movement through DRAM) could cut the dominant cost of GPU 3DGS. This observation is one of the key benefits of stripping away global memory.

\paragraph*{IPU vs PPAs}

A consequence of the NEWS routing scheme is that Gaussians travel one tile per BSP superstep, which is a latency that can be seen as analogous to data fetching from DRAM. This affects rendering from arbitrary, unrelated viewpoints, but is mostly hidden in the settings with slow viewpoint changes such as scrolling viewers or in online reconstruction. In these settings the camera moves only incrementally between frames, so the set of Gaussians projecting to each tile changes slowly and most primitives are already close to their correct destination. In the case of a GPU, the whole scene is repeatedly fetched from DRAM every new frame and requires caching to be efficient.

Random access through DRAM is only strictly necessary when consecutive views may come from arbitrary camera poses; for incremental view changes, local nearest-neighbour exchange is a natural fit. This is what PPAs exploit to achieve extremely high framerates. Local-only communication has proven performant for front-end vision algorithms (tracking, mapping, incremental reconstruction) and makes exploiting this in differentiable rendering an interesting avenue.

We are transparent about the gap between the IPU and actual pixel processor arrays. IPU tiles have 624\,KB of SRAM and can be programmed in C++; PPA pixels have bytes of storage and fixed-function ALUs. Bridging the gap may require:

\begin{itemize}
  \item \textbf{Compact representations}: reducing the 60-byte Gaussian struct to fit PPA-scale memory, perhaps via quantisation or learned compression~\cite{niedermayr2023compressed}.
  \item \textbf{Hierarchical approaches}: multi-resolution Gaussian representations where coarse levels fit on-sensor and fine detail is deferred.
  \item \textbf{Hybrid architectures}: combining PPA-style sensor-processors for front-end vision with more capable tiles for rendering.
\end{itemize}

Our work may also be applicable in other settings such as large scale rendering where scene data must be routed between a cluster of many CPUs. Since the IPU is a coarse-grain-reconfigurable-array (CGRA) the techniques we employ could be useful in a number of similar distributed systems such as cloud CPU clusters or Cerebras' Wafer-Scale-Engine \cite{Fouladipathtracingmanycpus, lie2023cerebras}. 

\vspace{-2mm}

\section{Analysis}
\label{sec:analysis}

We identify bottlenecks in our DRAM-free implementation for large scenes. As we do not implement any mitigations for these in our prototype, exceeding the preallocated buffer size may compromise both performance and render quality.
\vspace{-1mm}

\begin{figure}[t]
    \centering
    \includegraphics[width=\linewidth]{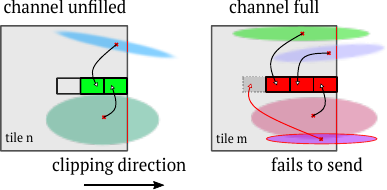}
  \caption{Inter-tile channel saturation. Each tile's outgoing NEWS channels (shown as black rectangles) have a fixed capacity set at compile time. When many Gaussians anchored in a dense region must simultaneously propagate to neighbouring tiles (overlapping to the right in the image), the outgoing buffer fills before all copies can be written (arrows), and excess Gaussians are dropped rather than forwarded. Tiles that should have received a copy are left without the Gaussian's contribution, producing rectangular holes aligned with tile boundaries in the rendered output (see \Cref{fig:tileArt}).}
    \label{fig:fullChannels}
\end{figure}

\paragraph*{1) Inter-tile bandwidth.}
The north-east-south-west (NEWS) channels between tiles have fixed capacity defined at compile time. In some cases, when many large Gaussians must be propagated to neighbouring tiles during the bloom phase, the outgoing channels saturate and some copies cannot be forwarded, as in \Cref{fig:fullChannels}. When a Gaussian that spans multiple tiles fails to reach one of them, that tile does not render its contribution, producing a visible hole aligned with the tile boundary. In dense regions where many Gaussians compete for limited channel capacity, these holes are noticeable as rectangular tiling artifacts (\Cref{fig:tileArt}).
\vspace{-1mm}

\paragraph*{2) Propagation latency.}
Both routing and bloom operate via local hops on the NEWS grid: a Gaussian can only advance one tile per BSP superstep, and each superstep corresponds to one rendered frame. A Gaussian whose anchor moves $d$ tiles after a viewpoint change therefore takes $d$ frames to arrive at its destination. During this transit the Gaussian is absent from the correct tile and may be rendered in the wrong location, producing a visible ``blooming'' effect where geometry appears to sweep across the screen before settling.
The artefact is visible for large camera movements and for Gaussians with large screen-space extent, which must bloom across many tiles after reaching their anchor.

\vspace{-1mm}

\paragraph*{3) Per-tile SRAM capacity.}
Each tile must buffer all incoming Gaussians in its 624\,KB SRAM.
After accounting for the tile's portion of the framebuffer, the codelet
binary and thread stacks, roughly 400\,KB remains for scene data.
We split this between eight inter-tile channels (one for each of the
four NEWS neighbours in each direction, $\sim$192\,KB total at a
capacity of 400 Gaussians per channel) and an internal storage buffer
($\sim$192\,KB, $\sim$3{,}200 Gaussians). Tiles in dense scene regions
can still exceed either budget, causing Gaussians to be dropped;
this produces the same holes and rectangular tiling artifacts from
\Cref{fig:tileArt}.

\paragraph*{4) Load imbalance.}
Gaussian density in screen space can be highly non-uniform. In the BSP model, all tiles synchronise at the barrier, so the slowest tile determines the superstep duration. Profiling reveals a high average percentage of execution time is spent waiting for the most loaded tiles to complete. We demonstrate this with the example in \Cref{fig:load_balancing} where the distribution of Gaussians across tiles changes with viewpoint: a close-up view distributes load relatively evenly, while a distant view concentrates all Gaussians onto a small number of central tiles. \vspace{1mm}

Offloading excess Gaussians to host DRAM, the strategy a GPU
manages in hardware, is feasible and would remove artifacts, but we
do not implement this to keep computation entirely on-chip. Beyond
this, several algorithmic mitigations would reduce the artifacts:
1) bloom only the highest-contribution Gaussians, 2) blend only high-opacity Gaussians, 3) train scenes to be more spatially uniform. We note that there are far fewer artifacts under continuous motion than
when teleporting, as most Gaussians are already near their anchor tile. We measure the routing substeps required to converge: a cold-start
teleport takes ${\sim}55$ substeps, whereas a small incremental change
(a $1.0\to1.2\times$ zoom) re-converges in only ${\sim}11$. Under the
incremental motion typical of SLAM and interactive viewing, the
per-frame routing cost stays low.

\vspace{-3mm}

\subsection{Compilation and Memory Allocation}

All these bottlenecks can potentially be alleviated by utilising the idle tiles and exploiting just-in-time compilation for arbitrary tile-to-tile reads at runtime. However, they are interesting to highlight as they can inform the trade-offs of algorithms and representations that better exploit spatial locality of data.  The channel/storage split presents an interesting trade-off when choosing buffer sizes: larger channels
let more Gaussians propagate per frame (reducing the bloom and
tile-boundary artifacts in \Cref{fig:tileArt}), but leave less
room for Gaussians already anchored to the tile, which reduces the detail that can be rendered there. Our current configuration allocates
$\sim$192\,KB to channels and storage; artifacts appear
as scenes approach this limit in dense regions, even when the mean
per-tile count is far lower.

\vspace{-3mm}

\subsection{Optimisation Opportunities}

The current implementation is a prototype with room for interesting optimisations:

\begin{itemize}
  \item \textbf{AMP engine}: IPU tiles contain a systolic array for matrix products that is not used in our hand-written codelets; exploiting it would accelerate the matrix operations such as projection.
  \item \textbf{Reduced precision}: We use 32-bit precision floating point for Gaussian parameters, this could be compressed to use lower precision as IPU supports FP8 and FP16 which would accelerate the algorithm and help store more Gaussians on-chip.
  \item \textbf{Exploiting MIMD execution}: Unlike GPUs and PPAs, IPU tiles can run entirely different codelets in the same superstep. Our current pipeline treats all tiles uniformly, similar to PPA-style SIMD constraints, but heterogeneous tile roles open interesting possibilities. For example, dedicating peripheral tiles to prefetching and compression while central tiles composite, or using idle tiles in distant-view configurations to run auxiliary work such as feature tracking alongside rendering.
  \item \textbf{Load balancing}: We could detect idle tiles at runtime and offloading projection work to them, exploiting the MIMD flexibility above to shift load dynamically rather than pinning each tile to a fixed screen-space region.
\end{itemize}

\vspace{-3mm}

\section{Conclusion}
\label{sec:conclusion}

We have presented the first implementation of 3D Gaussian Splatting's forward pass with only on-chip SRAM. We developed a NEWS grid routing scheme that distributes Gaussian primitives to tiles owning screen-space regions, entirely within the constraints of compile-time-defined communication.

Our implementation shows that 3DGS does not fundamentally require DRAM, once each Gaussian reaches the right tile, rendering is embarrassingly parallel, just as it is on the GPU. What changes is that data movement, normally hardware managed with a global address space and cache hierarchy, becomes a core part of the algorithm: DRAM access is replaced by a network of nearest-neighbour exchanges. Within that network we characterise the following bottlenecks: inter-tile bandwidth, per-tile SRAM capacity, and workload imbalance from non-uniform Gaussian density. Crucially, these bottlenecks are most acute when consecutive viewpoints are far apart, forcing Gaussians to traverse long paths.

A central theme of this work is the exploitation of spatial and
temporal locality. Standard GPU 3DGS re-sorts the entire Gaussian list in DRAM on every view change; in contrast, the fabric of the IPU allows us to move data between tiles only when necessary. The insight that motivates such a design is simple: when the viewpoint changes gradually, so does the screen-space projection of the scene, so most primitives remain where they were. Scrolling viewers, real-time game engines, and incremental path tracers all exploit spatial and temporal coherence, yet CUDA implementations of radiance fields do not, and treat
every frame as if the camera could teleport. For inverse-graphics
applications such as robotics and SLAM, the camera pose is grounded in real 3D motion and cannot jump arbitrarily, making such locality a
property we can reliably exploit.

Our experiments suggest that algorithmic choices that exploit data
locality pay off on any architecture where the cost of data movement dominates,
motivating further work on hierarchical and sparse connectivity
patterns. Another natural next step is the backward pass: gradients in an incremental setting, like the Gaussians in the forward pass,
would flow primarily between neighbouring tiles, so the same
locality argument carries over. In making data movement explicit, the hardware fosters decisions about placement, routing, and load balance of data that motivate further research in the area. We
see this as an early step toward rendering pipelines for on-sensor
and edge architectures, as well as toward GPU kernels that spend
less time waiting on DRAM.

\section{Acknowledgements}

This research was partly funded by the EPSRC On-Sensor Computer Vision grant (EP/Y020499/1). We would like to thank Riku Murai, Marwan Taher, Eric Dexheimer, and other members of the Robot Vision Group and Software Performance Optimisation Group for insightful discussions.

\bibliographystyle{eg-alpha-doi}
\bibliography{references}

\end{document}